\documentclass[preprint, aps]{revtex4}

\newcommand{\bq}{\begin{equation}}
\newcommand{\eq}{\end{equation}}
\newcommand{\bn}{\begin{eqnarray}}
\newcommand{\en}{\end{eqnarray}}

\begin{document}

\title{Electron Dynamics in a Coupled Quantum Point Contact Structure
with a Local Magnetic Moment \footnote{Based on work presented at
2004 IEEE NTC Quantum Device Technology Workshop}}
\author{Vadim I. Puller and Lev G. Mourokh }
\affiliation{Department of Physics and Engineering Physics,
Stevens Institute of Technology, Hoboken, New Jersey 07030, USA}
\author{A. Shailos and J. P. Bird}
\affiliation{Department of Electrical Engineering and Center for
Solid State Electronics Research, Arizona State University, Tempe,
Arizona, 85287-5706, USA}
\begin{abstract}
We develop a theoretical model for the description of electron
dynamics in  coupled quantum wires when the local magnetic moment
is formed in one of the wires. We employ a single-particle
Hamiltonian that takes account of the specific geometry of
potentials defining the structure as well as electron scattering
on the local magnetic moment. The equations for the wave functions
in both wires are derived and the approach for their solution is
discussed. We determine the transmission coefficient and
conductance of the wire having the local magnetic moment and show
that our description reproduces the experimentally observed
features.
\end{abstract}
 \maketitle

\section{Introduction}

The low-temperature conductance of quantum point contacts (QPCs)
is well known to be quantized in units of $2e^2/h$, a phenomenon
that can be explained in terms of a simple transmission (Landauer)
picture in which the influence of electron-electron interactions
is neglected \cite{Datta}. While this model is remarkably
successful in accounting for the observation of conductance steps
at integer units of $2e^2/h$, it is unable to explain the origin
of the additional conductance plateau, observed near $0.7\cdot
2e^2/h$ in numerous experiments. (For an overview of this issue,
see Ref. \cite{Science}.) While many different theoretical models
have been proposed to account for the origins of the 0.7 feature,
there is a wide consensus that it should be associated with some
novel many-body effect. In particular, there is growing consensus
that this feature is associated with the development of a net
magnetic moment in the QPC \cite{Meir1,Berggren,Hirose1}. In our
recent work \cite{Morimoto}, we have explored the use of coupled
quantum wires as a means for providing electrical detection of the
local-moment formation. The device structure that we have studied
is shown in Fig. 1 and was formed in the two-dimensional electron
gas of a GaAs/AlGaAs quantum well. The device  was realized by
means of electron-beam lithography, and lift-off of Ti-Au gates.
These gates were formed on a Hall bar with eight ohmic contacts,
positioned uniformly along its upper and lower edges. In suitable
combinations, these contacts could be used to make four-probe
measurements of the conductance of either wire, or of the quantum
dot itself (as indicated in Fig. 1). Of particular interest here
is the non-local measurement (right panel) that can be made by
measuring the conductance through one (fixed) wire as the gate
voltage ($V_g$) applied to the other (swept) wire is varied. The
key result of our experiment is that as the swept wire pinches
off, a resonant enhancement of the conductance of the fixed wire
is observed. A convincing theoretical explanation for this
resonant interaction was provided in a subsequent theoretical
report by our group \cite{Puller}. Based on a modified Anderson
Hamiltonian, we have shown that the resonant interaction with the
local magnetic moment formed in the swept wire leads to an
additional positive contribution to the density of states of the
fixed wire and, consequently, to an enhancement of its
conductance. While this analysis provides a qualitative
understanding of the resonant interaction between the quantum
wires, the tunnel matrix elements involved in the Anderson
Hamiltonian are generally unknown and have thus far been used as
fitting parameters. The influence of the specific device geometry
on these matrix elements has thus far been neglected, even though
geometry-related effects are known to be important for the
description of scattering in one-dimensional structures
\cite{Gurvitz1,Stone}. To overcome these shortcomings, in the
present paper we develop a more comprehensive theory for electron
dynamics in the coupled-wire system and attempt to calculate the
amplitude of the resonant inter-wire interaction from first
principles. The basic idea of this approach will be to calculate
the single-electron transmission properties in a device potential
that is modified by the presence of an extra scattering term,
arising from the presence of a local magnetic moment in one of the
wires. The formulation of this idea is given in Section II where
we derive equations describing the dynamics of electrons in the
swept and fixed wires. The approach to treat these equations is
given in Section III, where the expression for conductance is
obtained in terms of transmission matrix elements. In the present
paper we restrict ourselves to the analysis of electron dynamics
in the swept wire with the examination of the fixed wire to be
published elsewhere \cite{future}, and in Section IV we determine
the transmission coefficient and conductance for the swept wire.
In this, we obtain experimentally observed features such as
additional $0.75\cdot 2e^2/h$ plateau for the ferromagnetic
coupling and $0.25\cdot 2e^2/h$ plateau for the antiferromagnetic
coupling \cite{Rejec,Flambaum}. The conclusions are presented in
Section V.

\section{Electron modes in the coupled quantum wire structure}

We start our description of electron dynamics in the coupled
quantum wire structure from the following single-particle
Hamiltonian \bq
\hat{H}_{0}=K_{x}+K_{y}+U(x)+W(y)+V(x,y)-J(x,y)\hat{\vec{\sigma}}\cdot\hat{\vec{S}},
\label{eq:F1} \eq where $K_{x}$ and $K_{y}$ are the kinetic energy
operators for the electron localized in 2D plane, $W(y)$ is the
double-well potential describing the two quantum wires (Fig.2,
center panel), $V(x,y)$ is the potential of the tunnelling channel
connecting the two wires (Fig.2, right panel), and $U(x)$
describes the smooth bottleneck shape of the quantum wire
channels. The last term simulates exchange coupling between the
conductance electrons (Pauli matrices $\hat{\vec{\sigma}}$) and
the local moment, $\hat{\vec{S}}$, which is assumed to be a
spin-1/2 magnetic moment with $J(x,y) $ as the
coordinate-dependent exchange coupling constant. The potentials
$U(x),J(x,y),\,\textrm{and } V(x,y)\longrightarrow 0$ as
$x\longrightarrow \pm\infty$. The potential $V(x,y)$ is very sharp
in comparison with the variation of $U(x)$ in the $x$-direction.
$J(x,y)$ has an $x$-dependence similar to that of $U(x)$, since
the spatial characteristics of the local magnetic moment formed in
the conducting channel are determined by the shape of this
channel.

We write the Schr{\"o}dinger equation in the form \bq
\hat{H}_{0}\hat{\psi}(x,y)=E\hat{\psi}(x,y), \label{eq:F2} \eq
where the symbol "hat" in this and other equations is used for
operators and wave functions in the four-dimensional spin space of
the two spins. The basis vectors in this space are given by
\cite{KunzeBagwell} \bq
\hat{\chi}_{1}=\left|\uparrow_{e}\right\rangle
\left|\uparrow_{S}\right\rangle , \:
\hat{\chi}_{2}=\left|\downarrow_{e}\right\rangle
\left|\downarrow_{S}\right\rangle , %\nonumber\\
\hat{\chi}_{3}=\left|\uparrow_{e}\right\rangle
\left|\downarrow_{S}\right\rangle, \:\textrm{and }
\hat{\chi}_{4}=\left|\downarrow_{e}\right\rangle
\left|\uparrow_{S}\right\rangle ,\label{eq:F3} \eq where
$\left|\uparrow_{e}\right\rangle$
($\left|\downarrow_{e}\right\rangle$) and
$\left|\uparrow_{S}\right\rangle$
($\left|\downarrow_{S}\right\rangle$) are spin-up (spin-down)
states of the electron spin, $\vec{\sigma}$, and the local moment
spin, $\vec{S}$, respectively.

The solution of the Schr\"{o}dinger equation, Eq. (\ref{eq:F2}),
can be expanded in terms of the spin functions, Eq. (\ref{eq:F3}),
as
\begin{equation}
\hat{\psi}(x,y)=\sum_{\alpha=1}^{4}\chi_{\alpha}\psi_{\alpha}(x,y).
\end{equation}

It should be noted that the Hamiltonian, Eq. (\ref{eq:F1}), is not
diagonal in the spin space determined by the \emph{uncoupled}
representation, Eq. (\ref{eq:F3}), due to the presence of the
exchange term. This term can be diagonalized by means of a unitary
transformation to the \emph{coupled} representation with
transformation operator
\begin{equation}
\hat{X}=\left(%
\begin{array}{cccc}
  1 & 0 & 0 & 0 \\
  0 & 1 & 0 & 0 \\
  0 & 0 & \frac{1}{\sqrt{2}} & -\frac{1}{\sqrt{2}} \\
  0 & 0 & \frac{1}{\sqrt{2}} & \frac{1}{\sqrt{2}} \\
\end{array}
\right). \label{eq:Q3}
\end{equation}
The wave function in the coupled representation is given by
\begin{equation}
\hat{\psi}'(x,y)=\hat{X}^{+}\hat{\psi}(x,y) \label{eq:Q4}
\end{equation} with $\psi'_{\alpha} (\alpha =1,2,3)$ describing the
triplet and with $\psi'_4$ describing the singlet spin states. It
should be emphasized that for $x\longrightarrow \pm\infty$ the
Hamiltonian is diagonal in the uncoupled representation due to the
vanishing of potentials $U(x)=0$, $V(x,y)$ and $J(x,y)=0$.

Following the procedure of Ref. \cite{Gurvitz1} we expand the full
wave functions in terms of different propagating modes \bq
\hat{\psi}(x,y)=\sum_{n}\hat{\varphi}_{n}(x)\Phi_{n}(y) \eq  with
the transverse structure of n-th mode given by the solutions of
the equation \bq \left[K_{y}+W(y)\right]\Phi(y)=E_{n}\Phi_{n}(y).
\eq Correspondingly, the wave functions $\hat{\varphi}_{n}(x)$
obey the coupled equations \bq \left[
E-E_{n}-K_{x}-U_{n}(x)\right] \hat{\varphi}_{n}(x) =\sum_{m\neq
n}\left(V_{nm}(x)-J_{nm}(x)\hat{\vec{\sigma}}\cdot\hat{\vec{S}}\right)\hat{\varphi}_{m}(x)\label{eq:F7}
\eq where

 \bq V_{nm}(x)=\int dy \Phi_{n}^{*}(y)V(x,y)\Phi_{m}(y),
\eq \bq J_{nm}(x)=\int dy \Phi_{n}^{*}(y)V(x,y)\Phi_{m}(y),\eq and
$U_{n}=U(x)+V_{nn}(x)$.

In the following analysis we make a number of simplifications in
Eq. (\ref{eq:F7}). First, we note that if the wires are well
separated, the wave functions $\Phi_{n}(y)$ are strongly localized
in one of the two wires, therefore we can distinguish the modes
propagating in each of the wires. We assume that the shape of the
confining potential $W(y)$ is such that  one of the wires is close
to pinch off (\emph{swept} wire), i.e. it has only one propagating
mode (described by the wave function $\hat{\varphi}_{0}(x)$) with
the transverse confinement (subband bottom) energy, $E_{n}$, less
than the Fermi energy, whereas the other wire (\emph{fixed} wire)
has several propagating modes. The localized magnetic moment is
supposed to form in the only subband of the swept wire, hence the
exchange coupling can be approximated as
$J_{nm}(x)=\delta_{n,0}\delta_{m,0}J(x)$. Thus the system of
equations is reduced to

\bn \left[
E-E_{0}-K_{x}-U_{0}(x)+J(x)\hat{\vec{\sigma}}\cdot\hat{\vec{S}}\right]
\hat{\varphi}_{0}(x) = \sum_{n\geq
1}V_{0n}(x)\hat{\varphi}_{n}(x)\en  and \bq
\left[E-E_{n}-K_{x}-U_{n}(x)\right] \hat{\varphi}_{n}(x)
=\sum_{m}V_{nm}(x)\hat{\varphi}_{m}(x). \label{eq:F10} \eq

Furthermore, relying on the large energy separation between the
subbands in comparison with the magnitudes of $V_{nm}(x)$ and
$J(x)$, we restrict our analysis to a two-subband model, keeping
only the subband in the fixed wire whose energy is the closest to
that of the swept wire (the wave function of this subband is
$\hat{\varphi}_{1}(x)$) with the resulting set of equations given
by \bn \left[
E-E_{0}-K_{x}-U_{0}(x)+J(x)\hat{\vec{\sigma}}\cdot\hat{\vec{S}}\right]
\hat{\varphi}_{0}(x) = V(x)\hat{\varphi}_{1}(x), \en \bn
\left[E-E_{1}-K_{x}-U_{1}(x)\right] \hat{\varphi}_{1}(x)
=V(x)\hat{\varphi}_{0}(x),  \en \label{eq:F11} where we have
introduced $V(x)=V_{01}(x)=V_{10}(x)$.

Eqs. (14,15) can be decoupled using Green's functions: \bq
\hat{G}_{0}(\epsilon)=\left[\epsilon-K_{x}-U_{0}(x)+J(x)\hat{\vec{\sigma}}\cdot\hat{\vec{S}}\right]^{-1}
\label{eq:F12a} \eq and \bq
\hat{G}_{1}(\epsilon)=\left[\epsilon-K_{x}-U_{1}(x)\right]^{-1}.\label{eq:F12b}
\eq With these Green's functions Eqs. (14,15) can be formally
integrated as
 \bq
\hat{\varphi}_{0}(x)=\hat{G}_{0}(E-E_{0})V(x)\hat{\varphi}_{1}(x)
\nonumber \eq and \bq
\hat{\varphi}_{1}(x)=\hat{G}_{1}(E-E_{1})V(x)\hat{\varphi}_{0}(x).
\eq Accordingly, we obtain \bq \left[
E-E_{0}-K_{x}-U_{0}(x)+J(x)\hat{\vec{\sigma}}\cdot\hat{\vec{S}}\right]
\hat{\varphi}_{0}(x)
=V(x)\hat{G}_{1}(E-E_{1})V(x)\hat{\varphi}_{0}(x), \nonumber \eq
and \bq
 \left[E-E_{1}-K_{x}-U_{1}(x)\right] \hat{\varphi}_{1}(x)
=V(x)\hat{G}_{0}(E-E_{0})V(x)\hat{\varphi}_{1}(x).\label{eq:F13}
\eq

The Green's function $\hat{G}_{1}(\epsilon)$ is a scalar Green's
function, i.e. it is a unit matrix in the uncoupled spin space,
whereas $\hat{G}_{0}(\epsilon)$ has a more complicated structure.
Nevertheless, it can be expressed in terms of two scalar Green's
functions \cite{future} as \bq
\hat{G_0}(\epsilon)=\frac{1}{4}\left[3g^{t}(\epsilon)+g^{s}(\epsilon)\right]\hat{I}+
\frac{1}{4}\left[g^{t}(\epsilon)-g^{s}(\epsilon)\right]\hat{\vec{\sigma}}\cdot\hat{\vec{S}},\eq
where \bq g^{t}(\epsilon)=\left[ \epsilon
-K_{x}-U(x)+J(x)\right]^{-1} \label{eq:GG2a}\eq and \bq
g^{s}(\epsilon)=\left[ \epsilon
-K_{x}-U(x)-3J(x)\right]^{-1}.\label{eq:GG2b}
 \eq
Now we are able to redefine the scalar potentials, as \bq
\tilde{U}_{0}(x,E)=U_{0}(x)+V(x)\hat{G}_{1}(E-E_{1})V(x)
\eq\label{eq:F15} and \bq \tilde{U}_{1}(x,E)=U_{1}(x) +
V(x)\frac{1}{4}\left[3g^{t}(E-E_{0})+g^{s}(E-E_{0})\right]V(x),\label{eq:F16}
\eq and introduce the tunneling-induced exchange coupling of
electrons in the fixed wire to the local magnetic moment, \bq
j(x,E)=-V(x)\frac{1}{4}\left[g^{t}(E-E_{0})-g^{s}(E-E_{0})\right]V(x).\label{eq:F17}\eq
As a result, we obtain equations for the description of electron
dynamics in the swept and fixed wires in the form \bq \left[
E-E_{0}-K_{x}-\tilde{U}_{0}(x)+J(x)\hat{\vec{\sigma}}\cdot\hat{\vec{S}}\right]
\hat{\varphi}_{0}(x)=0 , \label{eq:F18a}\eq and \bq
\left[E-E_{1}-K_{x}-\tilde{U}_{1}(x)+j(x)\hat{\vec{\sigma}}\cdot\hat{\vec{S}}\right]
\hat{\varphi}_{1}(x) =0.\label{eq:F18b} \eq

Although the form of these two equations is very similar, and they
can be both treated in the same manner ( as is discussed in the
next Section), the results they yield will differ, depending on
the specific shapes of the potentials and exchange couplings. In
particular, while the shape of the coupling $J(x)$ in Eq.
(\ref{eq:F18a}) is smooth, similar to that of the potential
$U(x)$, the exchange constant $j(x)$ of Eq. (\ref{eq:F18b}) is
proportional to the potential $V(x)$, and therefore is sharper
than the bottleneck potential $U(x)$.

\section{Electron scattering by a localized spin}

In this Section we determine the transmission coefficient and,
correspondingly, the conductance of the quantum wire channel in
the presence of electron scattering from a magnetic moment. The
equation for the electron wave function in the uncoupled
representation has the form \cite{KunzeBagwell}
\begin{equation}
\left( \epsilon -K_{x}-U(x)+J(x)\vec{\sigma}\cdot \vec{S}
\right)\hat{\varphi}(x)=0. \label{eq:Q5}
\end{equation}
We are looking for a scattering solution of this equation with
incident wave of the form $\hat{A}e^{ikx}$, i.e. for a wave
incident from $x=-\infty$ and having momentum $\hbar
k=\sqrt{2m\epsilon}$. Here, $\hat{A}$ is also the vector in the
spin space, \bq
\hat{A}=\sum_{\alpha=1}^{4}A_{\alpha}\hat{\chi}_{\alpha},\eq with
$w_{\alpha}=\vert A_{\alpha}\vert^2$ being the probability to have
the certain initial orientation for the electron and magnetic
moment spins.

In the present paper we generalize the approach of Ref.
\cite{KunzeBagwell} (where the coupling constant $J(x)$ was
assumed to be a $\delta$-function) to the case of a real spatial
dependence of the coupling constant. To accomplish this, we find a
solution of Eq. (\ref{eq:Q5}) for the wave function in the coupled
representation and find the transmission coefficient for the wave
function in the uncoupled representation by means of unitary
transformation of Eq. (\ref{eq:Q3}). For the triplet and singlet
states of the coupled representation, Eq. (\ref{eq:Q5}) has the
form
 \bq \left( \epsilon
-K_{x}-U(x)+J(x)\right)\phi_{k}^{t\pm}(x)=0 \label{eq:Q9a}\eq and
\bq \left( \epsilon
-K_{x}-U(x)-3J(x)\right)\phi_{k}^{s\pm}(x)=0,\label{eq:Q9b}
 \eq
 where indices $t$ and
 $s$ denote the triplet and singlet solutions, respectively.
  The scattering solutions of these equations have the following asymptotic
  behavior \cite{Gurvitz1}
  \bq
\phi_{k}^{t,s+}(x)=\left\{\begin{array}{ll}
  T_{t,s}e^{ikx},\,& x\longrightarrow +\infty \\
  e^{ikx}+R_{t,s+}e^{-ikx},\,& x\longrightarrow -\infty \\
\end{array}\right.
\eq and  \bq \phi_{k}^{t,s-}(x)=\left\{\begin{array}{ll}
  e^{-ikx}+R_{t,s-}e^{ikx},\,& x\longrightarrow +\infty \\
  T_{t,s}e^{ikx},\,& x\longrightarrow -\infty \\
\end{array}\right. ,
\eq \label{eq:Q10} and the transmission coefficient of the wave
functions in the uncoupled representation can be obtained
\cite{future} as \bq \hat{T}=\left(
\begin{array}{cccc}
  T_{t} & 0 & 0 & 0 \\
  0 & T_{t} & 0 & 0 \\
  0 & 0 & \frac{1}{2}\left(T_{t}+T_{s}\right) &  \frac{1}{2}\left(T_{t}-T_{s}\right) \\
  0 & 0 &  \frac{1}{2}\left(T_{t}-T_{s}\right) &  \frac{1}{2}\left(T_{t}+T_{s}\right) \\
\end{array}
\right) \label{eq:Q18}\eq

Now we can use the Landauer-B\"{u}ttiker formula
\cite{Landauer,Buttiker} to determine the conductance of the
quantum wire: \bq
G=\frac{2e^2}{h}\sum_{\alpha,\beta}\left|T_{\alpha\beta}\right|w_{\beta}
\label{eq:Q19} \eq where $w_{\beta}$ gives the probability of
initial spin configuration. With the transmission matrix given by
Eq. (\ref{eq:Q18}), the conductance becomes \bq
G=\frac{2e^2}{h}\left[\left|T_{t}\right|^{2}\left(w_{1}+w_{2}\right)
+\frac{1}{2}\left(\left|T_{t}\right|^{2}+\left|T_{s}\right|^{2}\right)\left(w_{3}+w_{4}\right)\right]\label{eq:Q20}
\eq and, with account for the normalization of probability,
$\sum_{\alpha=1}^{4}w_{\alpha}=1$, it takes the form \bq
G=\frac{2e^2}{h}\left[\left|T_{t}\right|^{2}
+\frac{1}{2}\left(\left|T_{s}\right|^{2}-\left|T_{t}\right|^{2}\right)\left(w_{3}+w_{4}\right)\right]
. \label{eq:Q21} \eq

\section{Conductance of the swept wire} \label{sec:07}

 In this section we show how our model for the local magnetic moment reproduces the
 correct conductance behavior when the swept
 wire is pinched off. As the basis for our analysis we use the expression
 for the
 conductance, Eq. (\ref{eq:Q21}), jointly with our assumption that the potential
 ${U}_{0}(x)$ and the exchange coupling $J(x)$
 are smooth functions in comparison to the electron wavelength, $\lambda=2\pi/k$.
 It should be noticed that the potential $\tilde{U}_{0}(x)$  in Eq.
 (\ref{eq:F18a}) is not very smooth due to the additional
 contribution of a sharp potential $V(x)$ (see Eq. (23)),
 however our analysis is still valid as
 long as this additional contribution is small in comparison to
 $J(x)$.

 Functions $\tilde{U}_{0}(x)$ and $J(x)$ can be expanded into series near
 their maxima. Since the two functions are smooth we can assume that
  they take their maximum values at the same point $x=0$. These expansions
  are given by

 \bq
 \tilde{U}_{0}(x)=\tilde{U}_{0}(0)+\frac{x^2}{2}\frac{\partial^{2}\tilde{U}_{0}(x)}{\partial
 x^{2}}|_{x=0}=U_{max}-\frac{m\omega_{U}^{2}x^{2}}{2} \label{eq:SW1a}\eq and \bq
  J(x)=J(0)+\frac{x^2}{2}\frac{\partial^{2}J(x)}{\partial
 x^{2}}|_{x=0}=J_{max}-\frac{m\omega_{J}^{2}x^{2}}{2}.
 \label{eq:SW1b} \eq

Equations (\ref{eq:Q9a},\ref{eq:Q9b}), defining the transmission
coefficients $T_{t}$ and $T_{s}$, contain the effective potentials
$U_{+}(x)=\tilde{U}_{0}(x)-J(x)$ and
$U_{-}(x)=\tilde{U}_{0}(x)+3J(x)$. In view of Eqs.
(\ref{eq:SW1a},\ref{eq:SW1b}), these potentials can be treated as
inverse parabolic near their top, as \bq U_{+}(x)=U_{max}-J_{max}
-\frac{m\omega_{-}^{2}x^2}{2} \eq and \bq
U_{-}(x)=U_{max}+3J_{max} -\frac{m\omega_{+}^{2}x^2}{2}, \eq where
$\omega_{-}=\sqrt{\omega_{U}^{2}-\omega_{J}^{2}}$,
 $\omega_{+}=\sqrt{\omega_{U}^{2}+3\omega_{J}^{2}}$.

The transmission coefficient for the inverse parabolic barrier
$u(x)=-m\omega^{2}x^{2}/2$ is given by \cite{Levinson} \bq
t(\eta)=\left[1+e^{-2\pi\eta}\right]^{-1/2}, \eq where
$\eta=\epsilon/\hbar \omega$, and the energy, $\epsilon$, is
measured from the top of the barrier. Thus, for our situation we
obtain \bq T_{t}=t\left(\frac{\epsilon-U_{max}+J_{max}}{\hbar
\omega_{-}}\right)\eq and \bq
T_{s}=t\left(\frac{\epsilon-U_{max}-3J_{max}}{\hbar
\omega_{+}}\right) .\eq The most important feature of these
transmission coefficients is that the transmission probability,
$\left|t(\eta)\right|^{2}$, is very close to the step function.

Now we are able to calculate the conductance using Eq.
(\ref{eq:Q21}). We assume the equivalence of all initial spin
orientations, i.e. $w_{\alpha}=1/4$, and the conductance through
the swept wire takes the form \bn
G=\frac{2e^{2}}{h}\left[\frac{3}{4}\left|t\left(\frac{\epsilon-U_{max}+J_{max}}{\hbar
\omega_{-}}\right)\right|^2  %\nonumber \\
+\frac{1}{4}\left|t\left(\frac{\epsilon-U_{max}-3J_{max}}{\hbar
\omega_{+}}\right)\right|^{2}\right] . \en

The step-like structure of the transmission probability causes the
conductance to reproduce the step-like behavior of 0.7-anomaly. In
case of ferromagnetic coupling between the electrons and local
magnetic moment, $J_{max}>0$ our model gives an additional
conductance step at $0.75\times2e^{2}/h$, as  \bq
G=\frac{2e^{2}}{h}\left\{\begin{array}{ll}
  0,& \textrm{if } \epsilon<U_{max}-J_{max},  \\
  0.75, & \textrm{if } U_{max}-J_{max}<\epsilon<U_{max}+3J_{max}, \\
  1 ,& \textrm{if } \epsilon >U_{max}+3J_{max}. \\
\end{array}\right.
\eq It is interesting to point out that for antiferromagnetic
coupling , $J_{max}<0$, we obtain the a conductance step at
$0.25\times 2e^2/h$, which has been observed in experiments
 \cite{future} and density-functional simulations \cite{Berggren},
as
 \bq
G=\frac{2e^{2}}{h}\left\{\begin{array}{ll}
  0,& \textrm{if } \epsilon<U_{max}-3\left|J_{max}\right|,  \\
  0.25, & \textrm{if } U_{max}-3\left|J_{max}\right|<\epsilon<U_{max}+\left|J_{max}\right|, \\
  1 ,& \textrm{if } \epsilon >U_{max}+\left|J_{max}\right|. \\
\end{array}\right.
\eq

\section{Conclusion}

In conclusion, we have examined electron dynamics in  coupled
quantum wires  under conditions where a local magnetic moment is
formed in one of the wires. In our theoretical model, the
single-particle Hamiltonian has been employed with account of the
specific geometry defining the structure as well as electron
scattering on the local magnetic moment. We have derived the
coupled set of equations for the wave functions of electrons in
both wires and have been able to decouple them obtaining equations
describing electron dynamics in each of the two wires. While our
analysis of  electron processes in the fixed wire will be
published elsewhere \cite{future}, we determine the transmission
coefficient and conductance of the swept wire (having the local
magnetic moment) and show that our description reproduces the
experimentally observed features such as additional plateaus in
the conductance at $0.75\cdot 2e^2/h$ (for the ferromagnetic
coupling between electrons and the local magnetic moment) and at
$0.25\cdot 2e^2/h$ (in the case of the antiferromagnetic
coupling).

\newpage
\begin{figure}

  \caption{Electron micrographs of the critical region of the
  device,
indicating the two measurement configurations}\label{fig:1}
\end{figure}

\begin{figure}

  \caption{The figure on the left is a schematic of the
  experimental structure.   The confining potential associated
   with this structure is modeled as a sum of potentials defining
   the two wires, $W(y)$, (center) and defining the tunnel channel
   between them, $V(x, y)$, (right).}\label{fig:2}
\end{figure}

\end{document}